\documentclass[aps,prl,reprint,superscriptaddress,twocolumn,preprintnumbers]{revtex4-2}

\usepackage[utf8]{inputenc}\usepackage[T1]{fontenc}
\usepackage{amsmath}
\usepackage{amssymb}
\usepackage{graphicx}
\usepackage{slashed}
\usepackage{cases}
\usepackage[toc,page]{appendix}
\usepackage{empheq}
\usepackage[normalem]{ulem} 
\usepackage{xcolor} 
\usepackage{afterpage}
\usepackage{float}
\usepackage{bbold}
\usepackage{hypbmsec}
\usepackage{multirow}
\allowdisplaybreaks
\usepackage[export]{adjustbox}
\usepackage[colorlinks = true,
            linkcolor = blue,
            urlcolor  = blue,
            citecolor = blue,
            anchorcolor = blue]{hyperref}
\makeatletter
\newcommand\fake@math{}
\def\fake@math#1\){[math]}
\makeatother

\begin{document}

\preprint{FERMILAB-PUB-22-498-T}
\preprint{NUHEP-TH/22-06}

\title{A New Probe of Relic Neutrino Clustering using Cosmogenic Neutrinos}

\author{Vedran Brdar}
\email{vbrdar@fnal.gov}
    \affiliation{Theoretical Physics Department, Fermilab, P.O. Box 500, Batavia, IL 60510, USA}
     \affiliation{Northwestern University, Department of Physics \& Astronomy, 2145 Sheridan Road, Evanston, IL 60208, USA}
\author{P. S. Bhupal Dev}
    \email{bdev@wustl.edu}
    \affiliation{Department of Physics and McDonnell Center for the Space Sciences,  Washington University, 
    St. Louis, MO 63130, USA}
     \affiliation{Theoretical Physics Department, Fermilab, P.O. Box 500, Batavia, IL 60510, USA}
    \author{Ryan Plestid}
    \email{rpl225@g.uky.edu}
    \affiliation{Department of Physics and Astronomy, University of Kentucky, Lexington, KY 40506, USA}
    \affiliation{Theoretical Physics Department, Fermilab, P.O. Box 500, Batavia, IL 60510, USA}
    \author{Amarjit Soni}
    \email{adlersoni@gmail.com}
    \affiliation{Physics Department, Brookhaven National Laboratory, Upton, NY 11973, USA}

\begin{abstract}
We propose a new probe of cosmic relic neutrinos (C$\nu$B) using their resonant scattering against cosmogenic neutrinos. Depending on the lightest neutrino mass and the energy spectrum of the cosmogenic neutrino flux, a Standard Model vector meson (such as a hadronic $\rho$) resonance can be produced via $\nu\bar{\nu}$ annihilation. This leads to a distinct absorption feature in the cosmogenic neutrino flux at an energy solely determined by the meson mass and the neutrino mass, apart from redshift. By numerical coincidence, the position of the $\rho$-resonance overlaps with the originally predicted peak of the Greisen-Zatsepin-Kuzmin (GZK) neutrino flux, which offers an enhanced effect at higher redshifts. We show that this absorption feature in the GZK neutrino flux may be observable in future radio-based neutrino observatories, such as IceCube-Gen2 radio, provided there exists a large overdensity in the C$\nu$B distribution. This therefore provides a new probe of C$\nu$B clustering at large redshifts, complementary to the laboratory probes (such as KATRIN) at zero redshift.  
\end{abstract}
\maketitle 

\section{Introduction}
The detection of the cosmic neutrino background (C$\nu$B) is an extremely important problem in fundamental physics~\cite{Weinberg:1962zza}. So far, there has been only indirect evidence for C$\nu$B from precise measurements of the primordial elemental abundances in big bang nucleosynthesis (BBN)~\cite{Steigman:2012ve}, cosmic microwave background (CMB)~\cite{Planck:2018vyg},  and large-scale structure (LSS)~\cite{eBOSS:2020yzd, DES:2021wwk}.  However, the direct detection of  C$\nu$B remains an open challenge, which is often dubbed as the ``Holy Grail'' of neutrino physics. A direct detection of C$\nu$B will provide a strong validation of the Hot Big Bang cosmological model, and moreover, will provide a window into the first second of creation before the pre-recombination age. There exist several proposals for the direct detection of C$\nu$B~\cite{Gelmini:2004hg, 10.3389/fphy.2014.00030, Vogel:2015vfa, PTOLEMY:2018jst}, but none of them are expected to be experimentally feasible in the foreseeable future. 

Recall that in the standard cosmological picture, the effective temperature of the C$\nu$B today is inherently connected to the CMB temperature:
\begin{align}
    T_{\nu,0}= \left(\frac{4}{11}\right)^{1/3}T_{\gamma,0}=1.945~{\rm K}\simeq 1.7 \times 10^{-4}~{\rm eV} \, .
\end{align}
From neutrino oscillation data~\cite{Esteban:2020cvm}, we know two mass-squared differences, while the absolute value of the neutrino mass is still unknown. Assuming that the lightest neutrino mass $m_{\rm lightest}\gtrsim T_{\nu,0}$, the C$\nu$B today can be thought of as a non-relativistic gas of fermions with the number density of 
\begin{align}
    n_{\nu, 0} = \frac{3}{4}\frac{\zeta(3)}{\pi^2}g_\nu T_{\nu,0}^3 \simeq 56~{\rm cm}^{-3} 
    \label{eq:density}
\end{align}
per neutrino flavor (and similarly for antineutrinos). Here $g_\nu=1$ is the number of degrees of freedom for each neutrino. Although the C$\nu$B has the largest flux among all natural and man-made neutrino sources~\cite{Spiering:2012xe}, it is their small kinetic energy that makes any direct probe of C$\nu$B extremely challenging.   

An interesting idea for C$\nu$B detection is via its scattering off ultra high-energy (UHE) neutrinos or cosmic rays. In particular, it was pointed out long ago~\cite{Weiler:1982qy} (see also Refs.~\cite{Fargion:1997ft, Yoshida:1998it,Eberle:2004ua, Barenboim:2004di}) that resonant annihilation of C$\nu$B with UHE neutrinos can produce a Standard Model (SM) $Z$ boson on-shell (the so-called `$Z$-burst') for UHE neutrino energies of the order of
\begin{align}
    E_\nu =\frac{m_Z^2}{2m_\nu (1+z)} = \frac{(4.2\times 10^{22}~{\rm eV})}{(1+z)}\left(\frac{0.1~{\rm eV}}{m_\nu}\right) \, ,
    \label{eq:Z-burst}
\end{align}
where $z$ is the redshift at which the $\nu\bar{\nu}$ annihilation occurs in the sky. The cross-section for this process is pretty large: $\langle \sigma_{\nu\bar\nu}^{\rm ann.}\rangle=2\pi\sqrt 2 G_F\simeq 4\times 10^{-32}~{\rm cm}^2$, where $G_F$ is Fermi's constant. However, given the cosmological~\cite{Planck:2018vyg} and laboratory~\cite{KATRIN:2021uub} constraints on the neutrino mass which is required to be $\lesssim {\cal O}$(eV), the $Z$-burst energy~\eqref{eq:Z-burst} is clearly beyond the so-called GZK cut-off~\cite{Greisen:1966jv,Zatsepin:1966jv}. As a reminder, although the origin of UHE cosmic rays (UHECR) are unknown~\cite{Coleman:2022abf}, it is widely believed that the highest-energy neutrinos originate in the scattering of protons from the UHECR sources (with energy $E_p\geq 5.5\times 10^{19}$ eV) off the CMB photons, which resonantly produces $\Delta^+$ baryons (with $m_{\Delta}\approx 1232$ MeV) that subsequently decay to pions and nucleons, with neutrinos being one of the final decay products -- these are the so-called cosmogenic or GZK neutrinos~\cite{Berezinsky:1970xj, Stecker:1973sy, Hill:1983xs}.   Therefore, it is unlikely for any absorption features in the UHE neutrino spectrum due to the $Z$-burst to be observed, unless some super-GZK UHECR sources are established~\cite{Davoudiasl:2000hv,Csaki:2003ef,Anchordoqui:2003gm, Ringwald:2006ru, Lai:2006eh}. 
It is worth mentioning here that the resonance energy can be lowered to sub-GZK scale and even down to the PeV scale currently being probed  at neutrino telescopes such as IceCube, if the resonant state is sufficiently light. This has been discussed in the literature in the context of non-standard neutrino interactions with light mediators~\cite{Ioka:2014kca, Ng:2014pca,Ibe:2014pja,Blum:2014ewa, Araki:2014ona,Cherry:2014xra,Kamada:2015era, DiFranzo:2015qea,Cherry:2016jol,Altmannshofer:2016brv,Bustamante:2020mep,Creque-Sarbinowski:2020qhz,Carpio:2021jhu, Esteban:2021tub, Carpio:2022lqk}, however the Standard Model (SM) itself provides light vector mediators below the GeV scale that can mediate a resonance and therefore supplies a mechanism for sub-GZK absorption. 

\section{Vector-meson Resonances}
\begin{figure}[!t]
\includegraphics[width=0.25\textwidth]{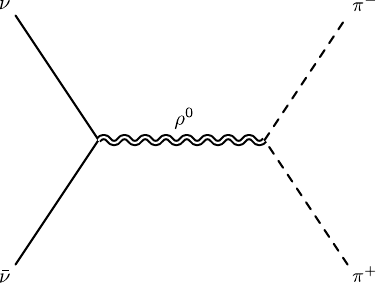}
\caption{Feynman diagram for the resonant production of neutral vector mesons (e.g., $\rho^0$) in $\nu\bar\nu$ annihilation, followed by their hadronic decay.} 
\label{fig:rho}
\end{figure}

Within the SM, even at much lower energies, there is the possibility of a vector (or axial-vector) meson resonance in $\nu\bar\nu$ annihilation.  Recall that in the SM, neutrinos and left-handed components of charged leptons are doublets of $SU(2)_L$: $\psi_i=(\nu_i, \ \ell_i^-)^T$, where $i=1,2,3$ is the family index. After electroweak symmetry breaking, the relevant piece of the SM Lagrangian reads~\cite{ParticleDataGroup:2020ssz} 
\begin{align}
    {\cal L} \supset -e Q_i \bar \psi_i \gamma^\mu \psi_i A_\mu - \frac{g}{2 \cos\theta_w} \bar \psi_i \gamma^\mu (g_V^i  - g_A^i \gamma_5 ) \psi_i Z_\mu \, , 
    \label{eq:SMLag}
\end{align}
where $A_\mu$ and $Z_\mu$ are the photon and the $Z$-boson fields respectively, $g$ is the $SU(2)_L$ gauge coupling, $\theta_w$ is the weak mixing angle,  $e=g\sin\theta_w$ is the positron electric charge, $Q_i$ is the fermion charge relative to the positron, and $g_{V,A}$ are the vector and axial-vector couplings for the $Z$-boson. 

Let us first recap the familiar case of $e^{+}e^{-}$ collisions. For the center-of-mass energy $\sqrt{s}\ll m_Z$, the $e^+e^-\to f\bar{f}$ ($f$ being any SM fermion) cross-section falls as $1/s$ for dimensional reasons (as long as $\sqrt s\gg m_f$). Therefore, the ratio $R$ of the cross-sections into quark-antiquark and $\mu^{+}\mu^{-}$ pairs is a constant that depends on the underlying number of quark degrees of freedom (see e.g.~Fig. 52.2 in Ref.~\cite{ParticleDataGroup:2020ssz}).  One can readily see that for $s\ll m_Z^2$, the $e^{+}e^{-}$ cross-section is dominated by the vector meson resonances with $J^{PC}=1^{--}$ (as predicted long ago in Refs.~\cite{Lee:1967iv, Gounaris:1968mw}), such as $\rho$, $\omega$ , $\phi$, $J/\Psi$,  $\Upsilon$ etc. --  an experimentally well-established fact.

In much the same way, when it comes  to $\nu \bar \nu$ collisions, the weak current contains an admixture of both vector and axial couplings [cf.~Eq.~\eqref{eq:SMLag}]. The weak vector current will produce the $J^{PC}=1^{--}$ resonances, such as the $\rho^0$ meson as shown in  Fig.~\ref{fig:rho}, while the axial-vector current will produce $J^{PC}=1^{++}$ resonances, such as $a_1(1260)$, $f_1(1285)$, $f_1(1420)$, $f_1(1510)$, $\chi_{c1} (3872)$,  $\chi_{b1}$, etc.~\cite{Bander:1994tc, Paschos:2002sj, Dev:2021tlo}. The corresponding cross-sections for both vector and axial-vector meson resonances can be calculated using the  Breit-Wigner resonance  formula~\cite{ParticleDataGroup:2020ssz}:
\begin{align}
  &  \sigma(\nu_i\bar{\nu}_i\to X^*) =  \frac{48\pi}{m_X^2} \frac{s\Gamma_X^2 {\rm BR}(X\to \nu_i\bar\nu_i)}{\left(s-m_X^2\right)^2+s^2\Gamma_X^2/m_X^2} \, ,
    \label{eq:crossSM}
\end{align}
where BR stands for the branching ratio of the resonance particle $X$ to neutrinos, $m_X$ and $\Gamma_X$ being its mass and total width respectively. The squared center-of-mass energy needed for a given resonance is $s=2m_\nu E_\nu$, where $E_{\nu}$ is the incident UHE neutrino energy: 
\begin{eqnarray}
E_{\nu} & = & 5\times 10^{18}~{\rm eV} \left(\frac{s}{{\rm GeV}^2}\right)\left(\frac{0.1~{\rm eV}}{m_\nu}\right) \, .
\end{eqnarray}
It is a numerical coincidence that the cosmogenic neutrino flux typically peaks around $10^{18}$~eV (depending on the exact value of the galactic-extragalactic crossover between the ``second knee'' and the ``ankle'' in the CR spectrum)~\cite{Ahlers:2010fw}, whereas the  lightest vector meson ($\rho^0$) resonance (with $m_\rho\approx 775$ MeV) requires a neutrino energy 
\begin{align}
    E_\nu = \frac{m_\rho^2}{2m_\nu (1+z)} \approx \frac{(3.0\times 10^{18}~{\rm eV})}{(1+z)}\left(\frac{0.1~{\rm eV}}{m_\nu}\right) \, .
    \label{eq:rho}
\end{align}
Therefore, we will mostly focus on the $\rho$-resonance in this work. Note that although the next lightest vector meson ($\omega$) mass ($m_\omega=782.7$ MeV) is close to the $\rho$-mass, its narrow width ($\Gamma_\omega=8.7$ MeV) makes a small difference to the broad resonance feature induced by the large $\rho$-width ($\Gamma_\rho\approx 150$ MeV). 
The $\rho$ production cross-section can then be estimated from Eq.~\eqref{eq:crossSM}, where the corresponding BR can be obtained from the partial width    
\begin{align}
   \Gamma(\rho\to \nu_i\bar\nu_i) & =  \frac{G_F^2}{24\pi}\left(1-2\sin^2\theta_w\right)^2f_\rho^2m_\rho^3 \, , 
\end{align}
where $f_\rho\approx 216$ MeV is the $\rho$-meson decay constant~\cite{Bharucha:2015bzk} (see also Refs.~\cite{Ebert:2006hj, Chang:2018aut}). We find that the cross-section at resonance is ${\cal O}(10^{-38}~{\rm cm}^2)$~\cite{Dev:2021tlo}, much smaller than the $Z$-burst cross section, primarily due to the hugely suppressed ${\rm BR}(\rho\to \nu_i\nu_i)\sim 10^{-13}$. Nevertheless, for charged current resonances e.g.\ $\bar{\nu}_e e^- \rightarrow \rho^-$ with a similar cross-section, ${\cal O}(100)$ events from the hadronic decay of the $\rho$-meson can   occur within the volume of IceCube for a 10-year exposure; whether these events can be detected is an open question~\cite{Brdar:2021hpy}. For the neutral current resonances like $\nu\bar{\nu}\to \rho^0$ under discussion, the corresponding event rates from $\rho$-decay are entirely negligible on Earth due to the low density of the C$\nu$B, i.e. $n_{\nu,0}\ll n_e$.  Instead, we propose to look for absorption features in the GZK neutrino spectrum itself due to the resonant production of the $\rho$-meson. As we show below, an observable effect requires an extremely large C$\nu$B overdensities at higher redshifts (i.e.\  $n_{\nu,z}/n_{\nu,0} \sim 10^{11}$) which we will assume to be the case. Possible origins of this overdensity will be discussed later.

\section{Attenuation of GZK Neutrinos}
As we discuss below, for an observable GZK absorption feature to appear in neutrino telescope data, it is necessary that GZK neutrinos have a high probability of encountering an overdense neutrino cloud. The most natural way for this to occur is if neutrino clustering is spatially correlated with UHECR production such that the GZK neutrinos are produced directly {\it within} the ultradense neutrino cloud. For an observable signature, it turns out to be necessary for UHECR production to be dominated by only a few sites in the Universe, and for a substantial fraction of the total neutrino population in one Hubble patch to be concentrated into clouds around these production sites. One may ask whether or not such a model is consistent with existing CR data. Due to the loss of directional information for charged CRs, due to e.g.\ deflection by galactic~\cite{Boulanger:2018zrk} or intergalactic~\cite{Garcia:2021cgu} magnetic fields, we argue that existing UHECR observations cannot rule out such a scenario. In fact, UHECR anisotropies have already been observed~\cite{PierreAuger:2017pzq, Abbasi:2021kjd}, and future measurements by large, ground-based experiments, like LHAASO~\cite{DiSciascio:2016rgi}, SWGO~\cite{Abreu:2019ahw} and IceCube-Gen2~\cite{IceCube-Gen2:2020qha} {\it can} in principle identify the origin of these anisotropies.
The scenario we propose can therefore be tested both by searching for the resonant absorption we discuss below {\it and} by the necessary anisotropy in the GZK neutrino flux. 

\begin{figure}[!t]
\includegraphics[width=0.5\textwidth]{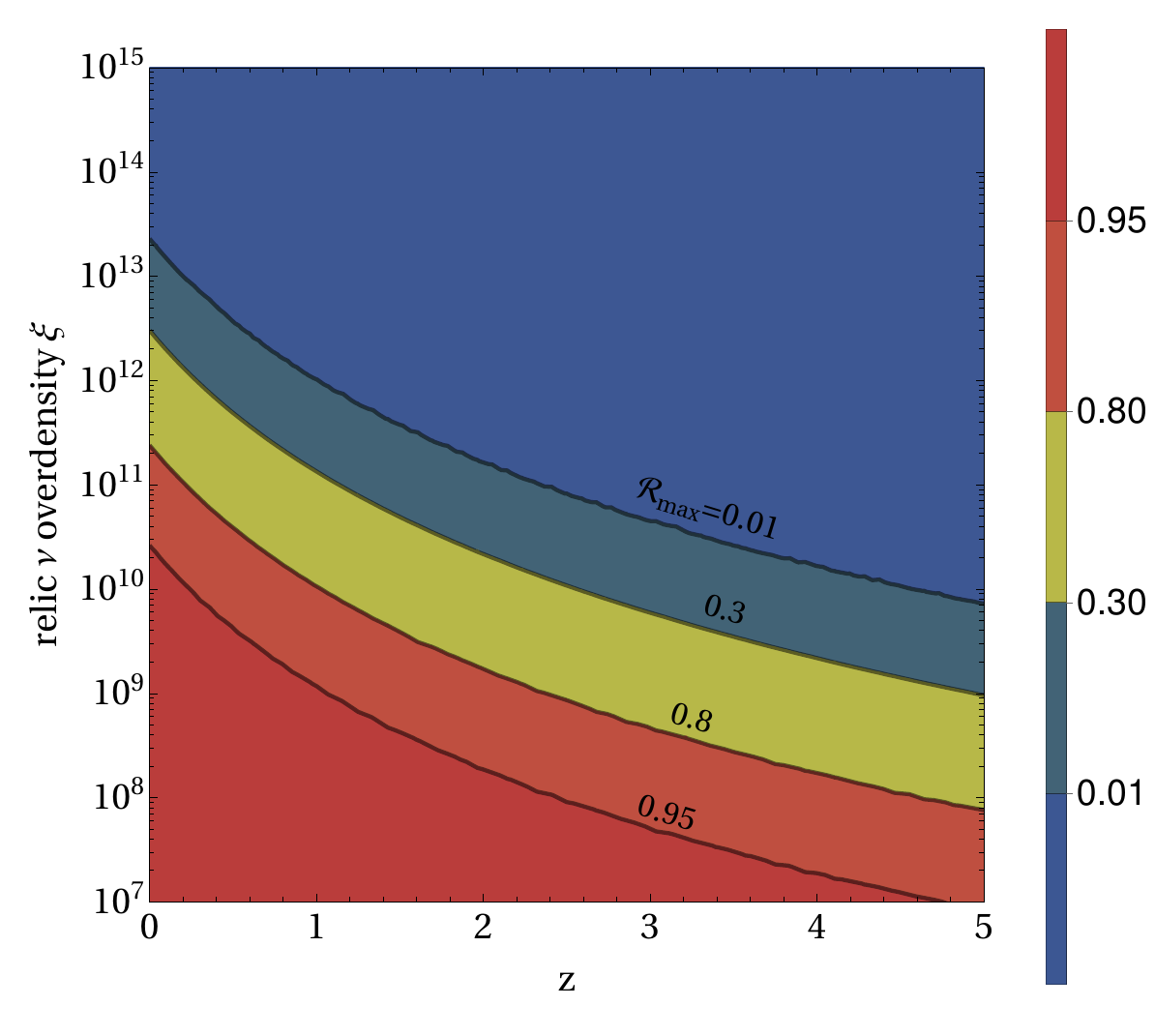}
\caption{The maximum attenuation factor for the GZK neutrino flux due to resonant $\rho$-meson production as a function of the redshift and C$\nu$B overdensity.} 
\label{fig:attn}
\end{figure}

Let us imagine, at some redshift $z$, the source of GZK neutrinos is surrounded by a cluster of C$\nu$B with density $n_{\nu}=\xi n_{\nu,0}(1+z)^3\equiv  n_{\nu,z}(1+z)^3$, where $n_{\nu,0}$ is given by Eq.~\eqref{eq:density} and $\xi$ quantifies the overdensity. The GZK neutrino flux will then be attenuated due to the resonant production of $\rho$-mesons, which promptly decay into pions and subsequently into charged leptons and neutrinos but typically with much lower energy than the original neutrino energy.  The attenuation factor is given by 
\begin{align}
    {\cal R} = e^{-L/\lambda} \, ,
    \label{eq:attn}
\end{align}
where $\lambda=1/\sigma n_\nu$ is the mean free path, $\sigma$ is the $\nu\bar{\nu}$ cross section given by Eq.~\eqref{eq:crossSM},  $L=(1/H_0)(q\xi)^{-1/3}$ is the average cluster length traversed by the GZK neutrinos (assuming a spherical cluster), $q$ is the total number of clusters in the Universe (assuming that each cluster is of the same size, and that the total number of relic neutrinos in the Universe is conserved), and $H_0$ the Hubble constant at $z=0$. The maximum attenuation at the resonant energy [cf.~Eq.~\eqref{eq:rho}] as a function of the redshift and overdensity is shown in Fig.~\ref{fig:attn}. It is clear that the attenuation effect is more pronounced at larger redshifts and/or larger overdensities simply because those two parameters directly impact the total C$\nu$B density at a given location. 

\begin{figure}[!t]
\includegraphics[width=0.5\textwidth]{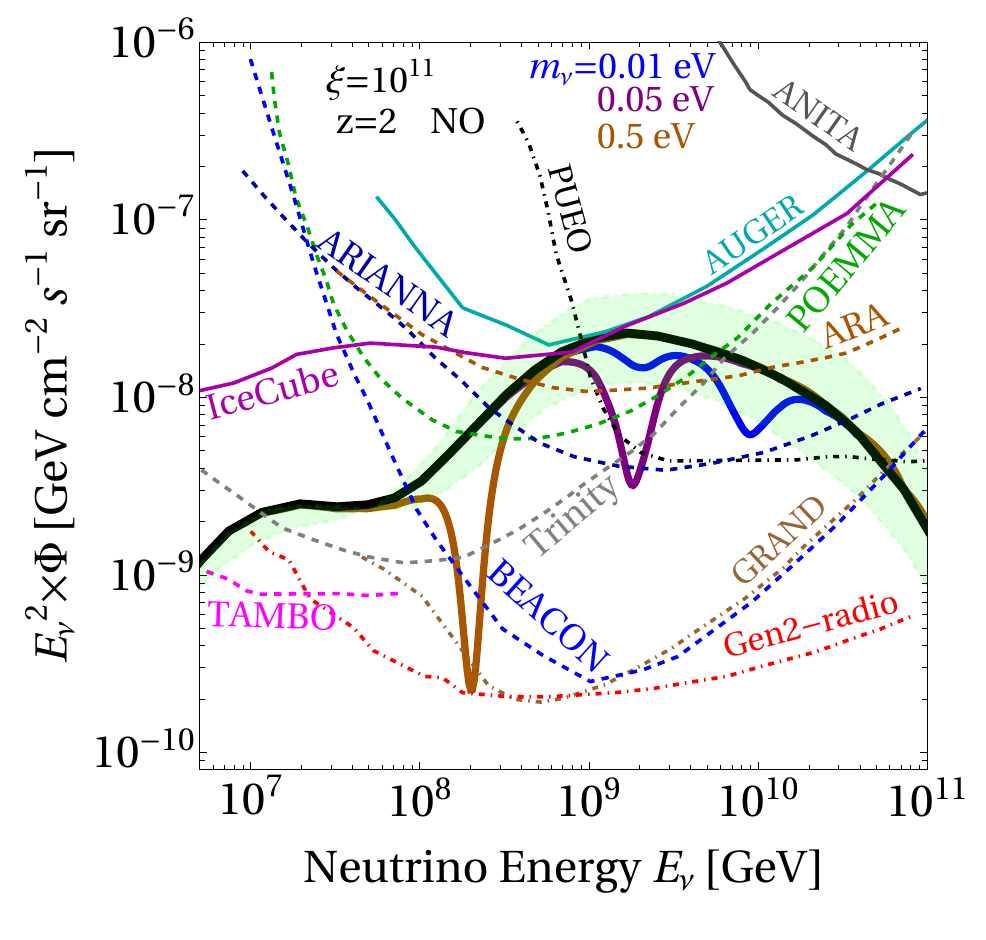}
\caption{Attenuation of the GZK neutrino flux, as compared to the unattenuated flux (solid black, with $99\%$ CL uncertainties in light green, from Ref.~\cite{Ahlers:2010fw}) due to resonant scattering with a $C\nu$B overdensity of $\xi=10^{11}$ at a redshift $z=2$ along the line of sight. We show the results for three benchmark values of the lightest neutrino mass $m_1=0.01$ eV (blue), 0.05 eV (purple) and 0.5 eV (orange). The normal ordering (NO) for neutrino masses is assumed. For comparison, the current constraints on the flux and future sensitivities (from Refs.~\cite{Ackermann:2022rqc,Huang:2021mki}) are also shown.} 
\label{fig:spectrum}
\end{figure}

Arbitrarily large redshifts are not realistic because the GZK neutrino flux, which is believed to originate from the progenitor cosmic-ray flux, should be correlated with the star-formation rate which peaks between $z=1$ and 2, and is essentially non-existent after $z=4$~\cite{Yuksel:2008cu}. Similarly, large overdensities are difficult to realize in concrete models, as we will discuss below, however they are not beyond the realm of possibilities for certain beyond the SM (BSM) scenarios. As a concrete example, we choose $z=2$, $\xi=10^{11}$ and $q=1$ in Eq.~\eqref{eq:attn} in order to illustrate the effect of resonant absorption on the GZK neutrino flux, as shown in Fig.~\ref{fig:spectrum}. Here the thick solid black curve is the unattenuated diffuse flux prediction from Ref.~\cite{Ahlers:2010fw} which is marginally consistent with present IceCube~\cite{IceCube:2018fhm} and Pierre Auger~\cite{PierreAuger:2019ens} upper limits; the green band corresponds to the CR models considered in Ref.~\cite{Ahlers:2010fw} with minimal and maximal energy density at the 99\% confidence level (CL). Note that this flux assumes primarily proton-dominated sources. Heavier elements like iron in the source composition  could in principle increase the flux uncertainty by an order of magnitude or even  more~\cite{Hooper:2004jc, Kotera:2010yn, Kampert:2012mx, Aloisio:2015ega, AlvesBatista:2018zui, Moller:2018isk, vanVliet:2019nse, Heinze:2019jou, Muzio:2021zud}. However, it is not clear which source composition is preferred by data, because the Pierre Auger data~\cite{PierreAuger:2019ens} with lower UHECR flux prefers heavier composition, whereas the TA data~\cite{TelescopeArray:2019mzl} prefers a higher flux. For a review, see e.g.,~Ref.~\cite{Kachelriess:2019oqu}.

 In Fig.~\ref{fig:spectrum}, the thick solid blue, purple and orange curves show the attenuated flux due to resonant $\rho$-production for three benchmark values of the lightest neutrino mass $m_1=0.01$ eV, 0.05 eV and 0.5 eV, respectively. Normal mass ordering is assumed, as seems to be moderately preferred by the latest oscillation data from T2K~\cite{T2K:2021xwb, Bronner:2022} and NOvA~\cite{NOvA:2021nfi, Hartnell:2022}. For $m_1=0.01$ eV, we can see two dips corresponding to the lighter mass eigenstates ($m_1, m_2$) combined and the heavy one ($m_3$), respectively. This is because the gauge couplings are flavor diagonal, and therefore, each mass eigenstate contributes separately to a single dip (unlike the case of secret neutrino interactions which can be flavor off-diagonal, and all mass eigenstates undergo absorption in all dips~\cite{Mazumdar:2020ibx, Esteban:2021tub}). But as the $m_1$ value increases, the three neutrino mass eigenstates become quasi-degenerate and the two peaks merge into one, as shown for the other benchmark points. 

Also shown in Fig.~\ref{fig:spectrum} by solid curves are the current constraints from ANITA~\cite{ANITA:2019wyx}, ARIANNA~\cite{Anker:2019rzo} and  ARA~\cite{ARA:2019wcf}, and by dashed/dotted curves are the future sensitivities of   GRAND~\cite{GRAND:2018iaj}, BEACON~\cite{Wissel:2020sec}, TAMBO~\cite{Romero-Wolf:2020pzh}, Trinity~\cite{Otte:2019knb}, POEMMA~\cite{POEMMA:2020ykm},  PUEO~\cite{PUEO:2020bnn} and  IceCube Gen2-radio~\cite{IceCube-Gen2:2020qha, IceCube-Gen2:2021rkf}. For more details, see Refs.~\cite{Ackermann:2022rqc, Huang:2021mki}. We find that the ``dip'' feature predicted here is within the sensitivity reach of some of these future experiments, such as GRAND, BEACON, Trinity and IceCube Gen2-radio.

\section{A Probe of Early C$\nu$B Overdensity} 
We now translate the result in Fig.~\ref{fig:spectrum} into an experimental sensitivity plot for the C$\nu$B overdensity as a function of the lightest neutrino mass, as illustrated in Fig.~\ref{fig:sens} for IceCube-Gen2 radio. Here we calculate the number of expected GZK neutrino events at IceCube-Gen2 radio with the unattenuated GZK flux (black curve in Fig.~\ref{fig:spectrum}) and compare it with the attenuated number of events for different values of the lightest neutrino mass assuming normal mass ordering. We have further assumed that forthcoming measurements 
of the GZK neutrino flux will converge to the presently unconstrained theoretical prediction shown in Fig.~\ref{fig:spectrum} and this is a realistic goal, provided the flux is primarily from proton-dominated CR sources. We perform a single-bin analysis by constructing an optimal bin size around the resonance energy for given values of lightest neutrino mass and redshift. If we take the full available energy range $E_\nu\in [10^{7},10^{11}]$ GeV (for which the effective area is known) as our  bin size, then we are simply performing a counting experiment in which a precise determination of the local flux attenuation in a narrow energy band is hindered. In the opposite limit, if we choose a bin width too narrow to capture the whole energy range corresponding to the width of the $\rho$ resonance, the sensitivity would also be poor. Therefore, we construct a bin width that matches the energy range corresponding to four widths of the $\rho$ meson (two on either side of the resonance energy where the bin is centered). We have checked that our results are stable against choosing slightly different bin widths, as well as against experimental energy resolution and smearing effects, which are expected to be at the level of ${\cal O}(10\%)$ (or better) for shower events -- so at least a factor of two better than the width-to-mass ratio of the resonance $\Gamma_\rho/m_\rho\simeq 19\%$. 

The number of unattenuated ($N_{\rm wo}$) and attenuated ($N_{\rm w}$) events is given by 
\begin{align}
    N_{\rm w/wo} =  \int_{(m_\rho-2\Gamma_\rho)^2/2m_\nu}^{(m_\rho+2\Gamma_\rho)^2/2m_\nu} dE \ T \ \Omega \ A_{{\rm eff}}(E) \ \Phi (E)  \ {\cal R}(E) \, , 
\end{align}
where $T$ is the exposure time (taken to be 10 years here), $\Omega=4\pi$ is the solid angle of coverage, $A_{\rm eff}$ is the IceCube-Gen2 radio effective area, $\Phi$ is the unattenuated GZK neutrino flux (cf. the solid black curve in Fig.~\ref{fig:spectrum}), and ${\cal R}$ is the attenuation factor given by Eq.~\eqref{eq:attn} (for the unattenuated case, ${\cal R}=1$). Then we compute the $\chi^2$ using the log likelihood method: 
\begin{align}
    \chi^2 = 2 \left(N_{{\rm w}}-N_{{\rm wo}}+N_{{\rm wo}}\log\frac{N_{{\rm wo}}}{N_{{\rm w}}}\right) \, .
\end{align}
The 90\% CL sensitivities ($\chi^2=2.71$ for one degree of freedom) for the IceCube-Gen2 radio with $T=10$ years of exposure are shown in Fig.~\ref{fig:sens} for two representative cases: (i) A single cluster of overdense neutrinos at $z=2$ (solid red curve); and (ii) A stochastic distribution of 10 neutrino clusters at $z=1$ (dashed red curve). 
For scenario (i), the sensitivity is clearly stronger because the cluster size is bigger [for $q=1$, the cluster length $L$ given below Eq.~\eqref{eq:attn} is of order ${\cal O}({\rm Mpc})$ for $\xi=10^{11}$]; therefore, the likelihood for resonant absorption increases. The bigger the number of clusters, the smaller their size (as we are restricted by the total number of C$\nu$B neutrinos in the Universe); therefore, we need a larger overdensity to see the absorption effect, as illustrated by the scenario (ii). The sensitivity in both cases is strongest for the lightest neutrino mass around $0.05$ eV because the corresponding resonant energy [cf.~Eq.~\eqref{eq:rho}] coincides with the peak of the predicted unattenuated GZK flux (see the purple benchmark in Fig.~\ref{fig:spectrum}). The flux goes down rapidly at higher energies which correspond to lower neutrino mass for resonance; therefore, the sensitivity curve goes up to the left. On the other hand, the IceCube-Gen2 radio effective area drops significantly for smaller energies, in particular below $E_\nu\sim 10^{17}$ eV;
for these energies, resonance is achieved for larger neutrino masses and this is why the sensitivity in Fig.~\ref{fig:sens} worsens for increasing neutrino mass. 

The solid purple curve in Fig.~\ref{fig:sens} is the current KATRIN upper limit on the overdensity: $\xi<1.1\times 10^{11}$ at 95\% CL,  derived using the possibility of C$\nu$B capture on tritium nuclei~\cite{KATRIN:2022kkv}. It is important to point out that the KATRIN limit is on the {\it local} overdensity, i.e. at $z=0$, whereas the IceCube-Gen2 radio limit derived here provides a {\it complementary probe} at higher redshifts. The vertical brown line in Fig.~\ref{fig:sens} is the KATRIN upper limit on the electron antineutrino mass: $m_{\bar\nu_e}< 0.8$ eV at 90\% CL~\cite{KATRIN:2021uub}, which is equivalent to an upper limit on each of the mass eigenvalues in the quasi-degenerate limit using the relation $m_{\bar\nu_e}^2=\sum_i|U_{ei}|^2m_i^2$ ($U$ being the PMNS mixing matrix~\cite{ParticleDataGroup:2020ssz}, assumed here to be unitary). Again, this bound is strictly applicable only at $z=0$. Note that neutrinoless double beta decay experiments have also imposed an upper limit on the effective neutrino mass $m_{\beta\beta} = |\sum_i U_{ei}^2 m_i|$ which  translates into a bound on $m_1\lesssim 0.2-0.6$ eV (depending on the nuclear matrix element used)~\cite{KamLAND-Zen:2016pfg, GERDA:2020xhi}; however, if neutrinos are Dirac particles, this bound does not apply, and therefore, is not included in Fig.~\ref{fig:sens}. 

Similarly, the vertical gray line in Fig.~\ref{fig:sens} is the 95\% CL Planck upper limit derived from the most stringent cosmological constraint on the sum of neutrino masses~\cite{Planck:2018vyg}. However, it is important to keep in mind that the this bound on $\sum m_i$ strongly depends on the combination of the cosmological datasets used, and varies from 0.12 eV to 0.60 eV (95\% CL)~\cite{Planck:2018vyg}. More importantly, the CMB bound is valid at a very high redshift of $z\approx 1100$, and there exists a number of ways to significantly relax the cosmological bound at lower redshifts, up to a few eV or so, by e.g.~assuming a non-standard cosmology~\cite{Farzan:2015pca, Bellomo:2016xhl,  Esteban:2021ozz, Esteban:2022rjk},  neutrinos with strong non-standard interactions ~\cite{Beacom:2004yd, Hannestad:2004qu, Chacko:2019nej, Escudero:2020ped, Barenboim:2020vrr}, neutrinos with a time-varying mass~\cite{Fardon:2003eh, Dvali:2016uhn, Ghalsasi:2016pcj, Lorenz:2018fzb, Lorenz:2021alz, deGouvea:2022dtw},  or neutrinos with a modified distribution function~\cite{Oldengott:2019lke, Alvey:2021sji}.          

Since neutrinos are fermions, they cannot be clustered to arbitrarily high densities in a stable configuration due to the Pauli exclusion principle~\cite{Tremaine:1979we}. Under the assumption that the relic neutrinos behave like an ideal Fermi gas, the maximum possible clustering depends on their Fermi energy $E_F$: $n_{\nu,0}^{\rm max} = \left(2m_\nu E_F\right)^{3/2}/(3\pi^2)$. 

 Using semi-analytic arguments which suggest an upper bound on the Fermi momentum $p_F\lesssim 0.9 \ m_\nu$, Ref.~\cite{Smirnov:2022sfo} obtained a maximum C$\nu$B density of $ n_{\nu,0}^{\rm max} \simeq 1.5\times 10^9~{\rm cm}^{-3} \left(m_\nu/0.1~{\rm eV}\right)^3$ for neutrino bound states (in presence of Yukawa interactions of neutrinos with a new light scalar). This theoretical bound is more stringent than the limits/sensitivities shown in Fig.~\ref{fig:sens}. 
Let us note, however, that since neutrino clustering at the scales we consider \emph{does} require BSM physics the phase space considerations are somewhat model dependent. Redshift dependent BSM effects, as well as the possible impact of non-standard cosmology on the evolution of the total neutrino energy density, or other exotic non-standard neutrino interactions could all conceivably modify the naive Fermi-gas phase space limit discussed above; we therefore, do not include phase space considerations in Fig.~\ref{fig:sens}.

\section{Discussion and Conclusion}
\begin{figure}[!t]
\includegraphics[width=0.5\textwidth]{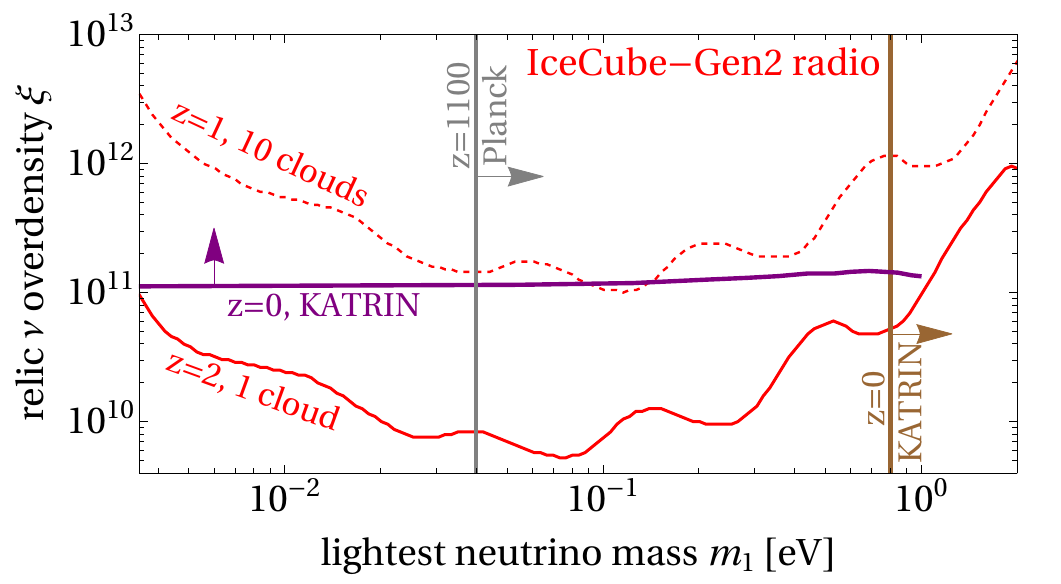}
\caption{IceCube-Gen2 radio sensitivity at 90\% CL to C$\nu$B overdensity as a function of the lightest neutrino mass. The red dashed and solid curves correspond to ($z=1,~q=10$) and ($z=2,~q=1$) respectively. The purple line is the current 95\% CL KATRIN upper limit on local overdensity~\cite{KATRIN:2022kkv}. The vertical brown line is the 90\% CL KATRIN upper limit on the neutrino mass~\cite{KATRIN:2021uub}. The vertical gray line is the 95\% CL Planck upper limit~\cite{Planck:2018vyg}.} 
\label{fig:sens}
\end{figure}

In principle, the resonant absorption mechanism proposed here is a purely SM phenomenon and does not require any BSM physics per se. However, as shown in Fig.~\ref{fig:sens}, the usefulness of this method in probing the allowed parameter space crucially depends on the existence of a large overdensity of C$\nu$B around or along the line of sight of the GZK neutrino source(s). As mentioned above, this is possible, which however most likely depends on new neutrino interactions with BSM fields. It is interesting to note that some of the proposals for evading the neutrino mass bounds also predict significant C$\nu$B clustering~\cite{Dvali:2016uhn, Dvali:2021uvk, Fardon:2003eh}. However, it still remains to be seen if ${\cal O}(10^{10})$ overdensities as needed in Fig.~\ref{fig:sens} can be achieved in practice. One might wonder if gravitational clustering around the GZK source could be of any help; but by itself, it can only provide overdensities up to $10^3$~\cite{Ringwald:2004np} (gravitational clustering at $z\gtrsim 0.5$ may be much larger than gravitational clustering around Earth at $z=0$). Another possibility is early neutrino decoupling; since the neutrino number density goes as $T^3$ [cf.~Eq.~\eqref{eq:density}], an earlier decoupling temperature by even a factor two in a non-standard cosmology could gain us an order of magnitude in the number density. A combination of several of these mechanisms may be necessary to produce the large overdensities that are experimentally accessible at near-term experiments.    

To conclude, we have proposed a new probe of relic neutrino overdensity which extends to higher redshifts. While the relic neutrino clustering will necessarily require BSM physics, the absorption mechanism relies exclusively on SM physics, namely, the scattering of relic neutrinos with cosmogenic neutrinos. Hadronic vector-resonances enhance the cross section and allow for neutrino telescopes to compete with bounds from KATRIN while simultaneously probing large redshifts. For large local neutrino densities that are correlated with the production sites of UHE cosmic rays, future radio-technology-based neutrino observatories like IceCube-Gen2 radio  can detect this ``dip'' feature. This provides a new probe of non-standard cosmologies beyond the current laboratory and cosmological bounds on neutrino masses.     

\section*{Acknowledgments}
We thank Carlos Arg\"{u}elles, Diego Aristizabal Sierra, Peter Denton, Abhish Dev, Francis Halzen, Shunsaku Horiuchi, Pedro Machado, Kohta Murase, Shmuel Nussinov, Jack Shergold and Ian Shoemaker for useful comments and discussions. In addition, we are grateful to Steffen Hallmann and Brian Clark for providing us the IceCube-Gen2 radio effective area. We are also grateful to Ibragim Alikhanov for pointing out a missing factor of $6$ in ~Eq.~\eqref{eq:crossSM} of v1, which now leads to slightly better results in ~Figs.  \ref{fig:spectrum} and \ref{fig:sens}. BD and RP acknowledge the Fermilab theory group for local hospitality during the completion of this work. RP also thanks the Weizmann Institute of Science for their hospitality. This work was partly performed at KITP, which is supported in part by the National Science Foundation under Grant No. NSF PHY-1748958, and at the Aspen Center for Physics, which is supported by the National
Science Foundation grant PHY-1607611. The work of BD was supported in part by the U.S. Department of Energy under Grant No. DE-SC0017987, by a Fermilab Intensity Frontier Fellowship, and by a URA VSP Fellowship. The work of RP was supported by the U.S. Department of Energy, Office of Science, Office of High Energy Physics, under Award Number DE-SC0019095. The work of AS was supported in part by the U.S. DOE contract \#DE-SC0012704.  This manuscript has been authored by Fermi Research Alliance, LLC under Contract No. DE-AC02-07CH11359 with the U.S. Department of Energy, Office of Science, Office of High Energy Physics.

\bibliographystyle{apsrev4-1}
\bibliography{references} 
\end{document}